\documentclass[%
reprint,amsmath,amssymb,aps,superscriptaddress  
]{revtex4-2}

\usepackage[utf8]{inputenc}
\usepackage[english]{babel}
\usepackage[T1]{fontenc}
\usepackage{amsmath}
\usepackage[colorlinks]{hyperref}
\usepackage{tikz}
\usepackage{lipsum}
\usepackage[normalem]{ulem} 

\definecolor{mygreen}{RGB}{50, 200, 50}

\usepackage{diagbox}
\usepackage[labelfont=bf]{caption}
\usepackage{multirow}
\usepackage{cleveref}

\makeatletter
\renewcommand\p@subsection{\thesection.}
\makeatother

\begin{document}

%
\title{Quantum computer specification for nuclear structure calculations}

\author{Ching-Hwa \surname{Wee}}
\affiliation{Department of Physics, Faculty of Science, Universiti Teknologi Malaysia, 81310 UTM Johor Bahru, Johor, Malaysia}

\author{Meng-Hock \surname{Koh}}
\email{kmhock@utm.my}
\affiliation{Department of Physics, Faculty of Science, Universiti Teknologi Malaysia, 81310 UTM Johor Bahru, Johor, Malaysia}
\affiliation{UTM Centre of Industrial and Applied Mathematics, Universiti Teknologi Malaysia, 81310 UTM Johor Bahru, Johor, Malaysia}

\author{Yung Szen \surname{Yap}}
\affiliation{Department of Physics, Faculty of Science, Universiti Teknologi Malaysia, 81310 UTM Johor Bahru, Johor, Malaysia}
\affiliation{Centre for Sustainable Nanomaterials (CSNano), Universiti Teknologi Malaysia, 81310 UTM Johor Bahru, Johor, Malaysia}
\affiliation{Centre for Quantum Technologies, National University of Singapore, 3 Science Drive 2, Singapore 117543, Singapore}

\begin{abstract}
Recent studies to solve nuclear structure problems using quantum computers rely on a 
quantum algorithm known as Variational Quantum Eigensolver (VQE).
In this study, we calculate the correlation energy in Helium-6 using VQE,
with a \textit{full-term} unitary-paired-coupled-cluster-doubles (UpCCD) ansatz on a quantum computer simulator and 
implement a set of custom termination criteria to shorten the optimization time.
Using this setup, we test out noisy quantum computer simulators of various coherence times and 
quantum errors to find the required specification for such calculations.
We also look into the contribution of errors from the quantum computers and optimization process.
We find that the minimal specification of 5~ms coherence times and $10^{-4}$ quantum errors is required to reliably reproduce state-vector results within 8\% discrepancy.
Our study indicates the possibility of performing VQE calculations using a full-term UpCCD ansatz on a slightly noisy quantum computer, 
without implementing quantum error correction.
\end{abstract}

\maketitle

Acknowledging the revolutionary potential brought by rapid advancement in quantum computing,
scientists from various fields of research have begun a concentrated effort
to incorporate aspects of quantum computing in their programme development.
One example is in the form of a white paper submitted to the U.S. Department of Energy in 2018~\cite{Carlson2018}
detailing a potential pilot program to develop quantum algorithms in the field of theoretical nuclear physics.

Within the nuclear physics community, increasing exploratory studies using quantum computers
to solve nuclear structure~\cite{DuWeiJie2021, Cervia2021, Stetcu2022, Romero2022, Siwach2022-pc, PrezFernndez2022, Perez-Obiol2023-vm, Du2024-gn, YangHongLi2023, Gibbs2024-gn, Yang2023-vd, Bhoy2024} and nuclear reaction~\cite{Roggero2020-sr, Du2021, Li2022, Turro2023, Rethinasamy2024-tm, Wang2024-xj} problems have been performed.
Many of these studies have been done on an ideal simulated quantum computer using state-vector simulation~\cite{Chikaoka2022, Qian2022-om, Romero2022, Kiss2022, Perez-Obiol2023-vm, Sarma2023, Bhoy2024}. 
In cases where a noisy quantum computer is used,
one often relies on a simplified ansatz with the purpose of 
shortening the quantum circuits depth~\cite{Dumitrescu2018, Kiss2022, Qian2022-om, Sarma2023}.
The reason being the short coherence times of the qubits, and the high error rates of the quantum gates
which limits the application of current Noisy Intermediate-Scale Quantum (NISQ) quantum computers.
Nevertheless, one would expect that this limitation would be lifted with further advancement in quantum computers.

\begin{figure}[hbt!]
    \centering
    \includegraphics[width = 0.9\linewidth]{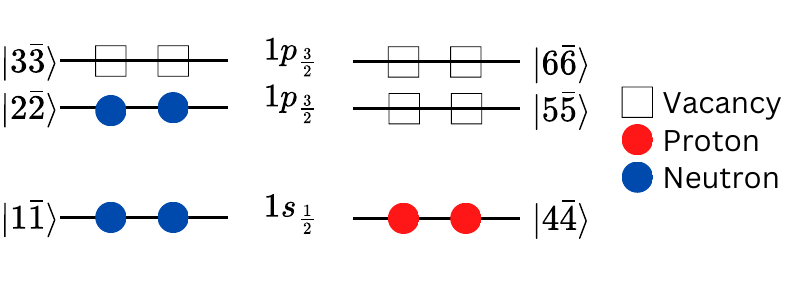}
    \caption{Bound energy levels in $^6$He. The figure shows the Hartree-Fock reference states, where all nucleons occupy the lowest energy levels. The occupied states are denoted by 
    $\left| i \bar{i}\right>$ where $i\in\left\{1,2,4\right\}$, 
    and the vacant states are denoted by $\left| j \bar{j}\right>$ where $j\in\left\{3,5,6\right\}$.
    The indices $\bar{i}$ and $\bar{j}$ refers to the time-conjugate of the corresponding
    $i$ and $j$ states, respectively.}
    \label{fig: energy_levels}
\end{figure}

Meanwhile, we navigate the challenges of current quantum computers by investigating the necessary
specifications to solve an actual nuclear physics problem related to nuclear pairing using  
Variational Quantum Eigensolver (VQE)~\cite{Peruzzo2014, Tilly2022}, 
which minimizes the expectation value of an observable with respect to an ansatz.
In our work, we use the full-term ansatz~\footnote{Full-term ansatz herein refers to an ansatz
that is sufficiently expressive and captures all essential features, e.g.~the symmetry requirement of the system under study.
}
instead of simplified ones.
We limit ourselves to solve a small system of $^6_2$He nucleus with only two neutrons on top of the magic $^4_2$He nucleus (Figure~\ref{fig: energy_levels}).

We relied on IBM 16-qubit Guadalupe superconducting quantum computer simulator, known as FakeGuadalupe (see Table~\ref{tab: FakeGuadalupe spec}) for the investigation.
While keeping most properties of the FakeGuadalupe, we modify (see Methods~\ref{subsec: FakeJohors spec}):
\begin{enumerate}
\item the qubit thermal relaxation ($T_1$) and qubit dephasing time ($T_2$)~\cite{Clarke2008} -- collectively termed here as coherence times,
\item quantum gate errors, readout errors, and ``state preparation and measurement (SPAM) errors''~\cite{Wood2020}
-- collectively referred here as quantum errors.
\end{enumerate}

\begin{table}[h!]
\caption{Specifications of FakeGuadalupe. Breakdown of coherence times into $T_1$ relaxation time and $T_2$ dephasing time; and quantum errors into quantum gate errors (1QGate and 2QGate refer to one-qubit and two-qubit gate errors), readout errors and SPAM errors.}

\def\arraystretch{1.3}
\setlength\tabcolsep{10pt}
\begin{tabular}{|ccccc|}
\hline
\multicolumn{1}{|c|}{}       & \multicolumn{1}{c|}{min}   & \multicolumn{1}{c|}{max}   & \multicolumn{1}{c|}{mean}  & std   \\ \hline
\multicolumn{5}{|c|}{Coherence Times (ms)}                                                                                  \\ \hline
\multicolumn{1}{|c|}{$T_1$}     & \multicolumn{1}{c|}{0.039} & \multicolumn{1}{c|}{0.119} & \multicolumn{1}{c|}{0.070} & 0.022 \\ 
\multicolumn{1}{|c|}{$T_2$}     & \multicolumn{1}{c|}{0.015} & \multicolumn{1}{c|}{0.142} & \multicolumn{1}{c|}{0.088} & 0.029 \\ \hline
\multicolumn{5}{|c|}{Quantum Errors ($\times10^{-2}$)}                                                                                        \\ \hline
\multicolumn{1}{|c|}{1QGate} & \multicolumn{1}{c|}{0.00}  & \multicolumn{1}{c|}{0.18}  & \multicolumn{1}{c|}{3.03}  & 3.56  \\ 
\multicolumn{1}{|c|}{2QGate} & \multicolumn{1}{c|}{0.68}  & \multicolumn{1}{c|}{1.99}  & \multicolumn{1}{c|}{1.08}  & 3.53  \\ 
\multicolumn{1}{|c|}{Readout} & \multicolumn{1}{c|}{1.06} & \multicolumn{1}{c|}{6.05} & \multicolumn{1}{c|}{1.98} & 1.20 \\ 
\multicolumn{1}{|c|}{SPAM}   & \multicolumn{1}{c|}{0.16}  & \multicolumn{1}{c|}{9.12}  & \multicolumn{1}{c|}{1.98}  & 1.78  \\ \hline
\end{tabular}%

\label{tab: FakeGuadalupe spec}
\end{table}

In this study, we have defined  a set of termination criteria based on a successive fit of data to a logarithmic line.
We have shown that such an approach yielded similar results to what would be obtained with a much larger maximum iteration.
Furthermore, we found that calculation errors originating from quantum computer specifications 
overweight the errors from optimization process.
Finally, our results indicate that efforts to extent the coherence times are more crucial than
the reduction of quantum errors since
our estimated quantum errors is currently of the same order with the state-of-the-art experimental achievement.

\begin{figure*}[!hbt]
    \centering
    \includegraphics[width = 1.0\textwidth]{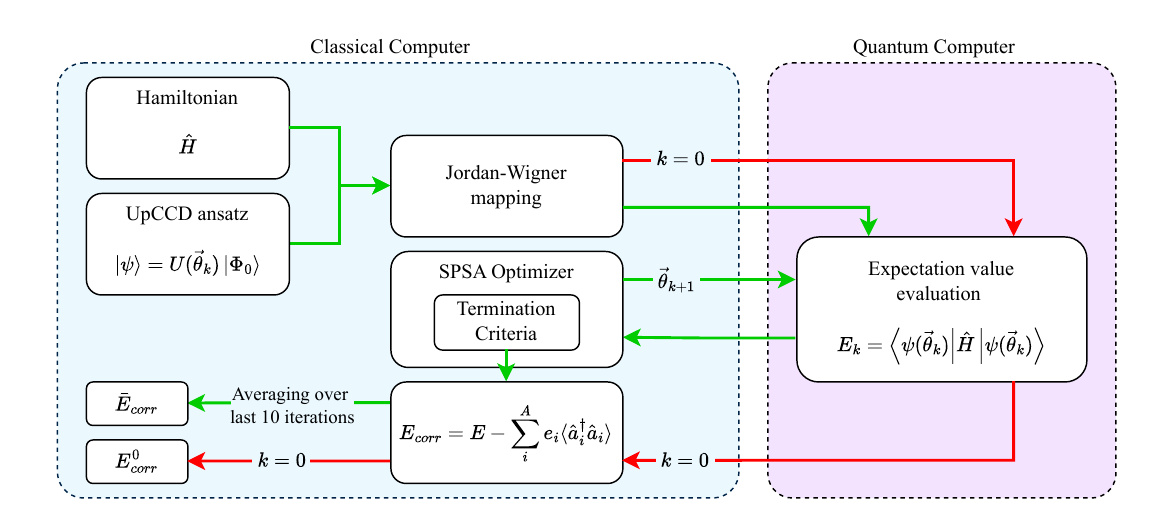}
    \caption{Framework of this study.
    The calculation begins with definition of a Hamiltonian, $\hat{H}$, and a UpCCD ansatz, $\left| \psi \right>$ starting from $\left| \Phi_0\right>$ initial state (Section~\ref{sec: NP on QC}).
    An expectation value of $\hat{H}$ with respect to $\left| \psi \right>$ is constructed (Methods~\ref{subsec: circuit construction}), then mapped into quantum circuits using Jordan-Wigner mapping, and evaluated using a quantum computer simulator.
    The SPSA optimizer (Methods~\ref{subsec: SPSA optimization}) then uses the expectation value to update the parameters for the next iteration $\vec{\theta}_{k+1}$, where $k$ represent current iteration number.
    When termination criteria are satisfied (Section~\ref{sec:convergence criteria}), $E_{corr}$ is extracted and $\bar{E}_{corr}$ is obtained by taking an average of the last 10 iterations.
    The green arrows show the process to obtain the $\bar{E}_{corr}$ to gauge performance of FakeJohors (Section~\ref{sec: fakeJohors performance}); this involves using simulated quantum computers of different specifications (Methods~\ref{subsec: FakeJohors spec}). The red arrows show the process of obtaining the correlation energy evaluated at $k=0$, labelled as $E^0_{corr}$, which does not go through optimization process (Section~\ref{sec: opt vs meas}).}
    \label{fig: framework}
\end{figure*}

\section{Calculations of nuclear correlation energy on quantum computer}\label{sec: NP on QC}
The Hamiltonian to be solved is given by
\begin{equation}
    \hat{H} =\sum_{i}e_{i}\hat{a}_{i}^{\dagger }\hat{a}_{i} + 
        \frac{1}{2}\sum\limits_{ij}V_{i\bar{i} j \bar{j}}\hat{a}_{i}^{\dagger}\hat{a}_{\bar{i}}^{\dagger }\hat{a}_{\bar{j}}\hat{a}_{j},
    \label{eq:Hamiltonian}
\end{equation}
where $e_{i}$ is the Hartree-Fock single-particle energy obtained using the Skyrme SLy4~\cite{Beiner1975} parametrization,
$V_{i\bar{i}j\bar{j}}$ is antisymmetrized pairing matrix element, 
$\hat{a}^\dagger$ and $\hat{a}$ are the fermionic creation and annihilation operators
for the single-particle state $i$ and its conjugate state $\bar{i}$.

A constant pairing matrix element \cite{Bonche1989} is used such that 
\begin{equation}
    V_{i\bar{i} j \bar{j}}^q = \frac{G_{q}}{11 + N_{q}}
    \label{eq:constant pairing}
\end{equation}
where $N_{q}$ represents the nucleon number of charge state $q \equiv \{n,p\}$, 
and $G_q$ represents the pairing intensity for both neutrons and protons chosen to be 1~MeV. 
Using pairing intensity of 1~MeV, Hartree-Fock--plus--Bardeen-Cooper-Schrieffer (HF+BCS) calculation 
assuming only \textit{like}-nucleons~\footnote{Nucleons of the same charge state e.g.~neutron and neutron. 
However, the two protons does not contribute to pairing since number two is a nuclear magic number in which the physics of the system is well reproduced using the independent particle framework. 
The neutron pairing while non-vanishing is rather minimal with BCS pairing gap amounting to only 0.07~MeV.
However, this choice of $G_n = 1$~MeV is necessary to achieve a convergence for a constrained solution at sphericity.}
pairing yielded a binding energy of \mbox{-29.28~MeV} as compared to experimental data of \mbox{-29.27~MeV} \cite{NNDC}.

The minimal pairing contribution in the case of $^6$He nucleus was chosen
as it provides a stringent test
on the capability of a quantum computer in reproducing a quantity at the order of about 0.1~MeV.
Using the single-particle energy levels from the HF+BCS calculations
and the pairing matrix elements generated using $G_n = G_p = 1$~MeV,
we determine the correlation energy, $E_{corr}$, defined as
\begin{equation}
    E_{corr} = E - \sum_i^{A}   e_i \langle\hat{a}^{\dagger}_i \hat{a}_i \rangle
    \label{eq: correlation energy}
\end{equation}
where $E$ is the expectation value of the Hamiltonian while
the summation of single-particle energy, $e_i$, involves the lowest occupied levels.

We solve the Hamiltonian within the unitary coupled cluster (UCC) framework, readily available in the 
Qiskit library~\cite{Qiskit} (see Methods~\ref{subsec: circuit construction}).
To account for pairing, we further limit to a unitary paired coupled cluster double excitations (UpCCD) ansatz~\cite{Henderson2014, Lee2018}, where only excitation of nucleons to paired-states on top of the Hartree-Fock reference state $\left\vert\Phi_{0}\right\rangle$ were considered.

For implementation on a quantum computer, we employed the Jordan-Wigner mapping~\cite{Tranter2018} to encode the fermionic operators into Pauli operators. 
Only the first order trotterized form was considered for the UpCCD ansatz.
While other types of mapping are available, e.g.~Parity~\cite{Seeley2012-fx} and Bravyi-Kitaev~\cite{Bravyi2002-qh} mapping, we chose the Jordan-Wigner mapping due to its intuitive representation where each single-particle level is mapped to only one qubit.

The parameters in the UpCCD ansatz have to be optimized to yield the lowest-energy solution. 
This was performed using
the simultaneous perturbation stochastic approximation (SPSA)~\cite{Spall1992-fm} 
because of its ability in handling noisy optimization~\cite{Spall1998, Radac2011, Finck2011}. 

For a better representation of the connectivity between various components discussed above,
readers are suggested to refer to the research framework
shown in Figure~\ref{fig: framework}.
The VQE approach, as well known, utilizes classical computer for majority part of the process.
The quantum computer (simulator in our case) is employed only for 
the evaluation of the expectation value of the Hamiltonian.

It is important to note that unlike classical supercomputers, comparison between different quantum computers is not straight forward due to differences in qubit layout/connectivity, coherence times, quantum error rates, supported native gates and consequently the depth of the quantum circuit, as also reported by Ref.~\cite{Buonaiuto2024-vq}.
Our approach herein, is then to modify only the coherence times and quantum errors for a specific quantum computer design chosen here to be the FakeGuadalupe.
The modified specifications of FakeGuadalupe is referred herein as FakeJohors.

\section{Termination criteria for faster convergence} \label{sec:convergence criteria}
Within the SPSA optimization process, it is customary that calculations
are terminated only at the maximum iteration.
This, however, consumes significant amount of time and computational resources.
To navigate this issue, we introduce a set of termination criteria,
which would allow the calculations to be terminated upon reaching a pre-defined criteria.

The termination criteria are based on a fit of measured value using a logarithmic equation
\begin{equation}
    y = m_k \ln{x} + c_k,
    \label{eq: fitted log line}
\end{equation}
at specific optimization step chosen such that $k = 10, 20, 30, \cdots, 200$.
The logarithmic fit takes into account all the preceding data up to the specific $k^{th}$ step.

The calculations will terminate whenever \textit{any} of these three criteria is triggered:
\begin{itemize}
    \item Criteria 1: $m_k>0$.
    \item Criteria 2: $\left\|m_{10}\right\| < 0.1$
    \item Criteria 3: $\Big| \frac{m_{k} - m_{\left(k-10\right)}}{m_{\left(k-10\right)}} \Big| \leq 8\%$.
\end{itemize}

The first criterion ensures that successive iterations lead to a lower energy,
and rules out cases where the calculated energy is higher than the Hartree-Fock solution.
In cases where the decrement in energy from $k = 1$ to $k = 10$ is rather small 
(which we define here by the absolute value to be at least greater than 0.1),
calculations will be terminated and repeated.
Such situation may occur, for example, when one is in stuck in a barren plateau~\cite{Larocca2024-be}.
Finally, the third criterion defines a way to properly terminate the calculations
by comparing the change in the slope $m$ between successive logarithmic fits
(e.g.~between $m_{20}$ and $m_{10}$ or between $m_{60}$ and $m_{50}$); 
the calculation is terminated when the change is less than 8\%.
When convergence is achieved, we determine the correlation energy by averaging the data over the last 10 iterations and denote this as $\bar{E}_{corr}$.

\begin{figure*}[hbt!]
    \includegraphics[width=0.95\textwidth]{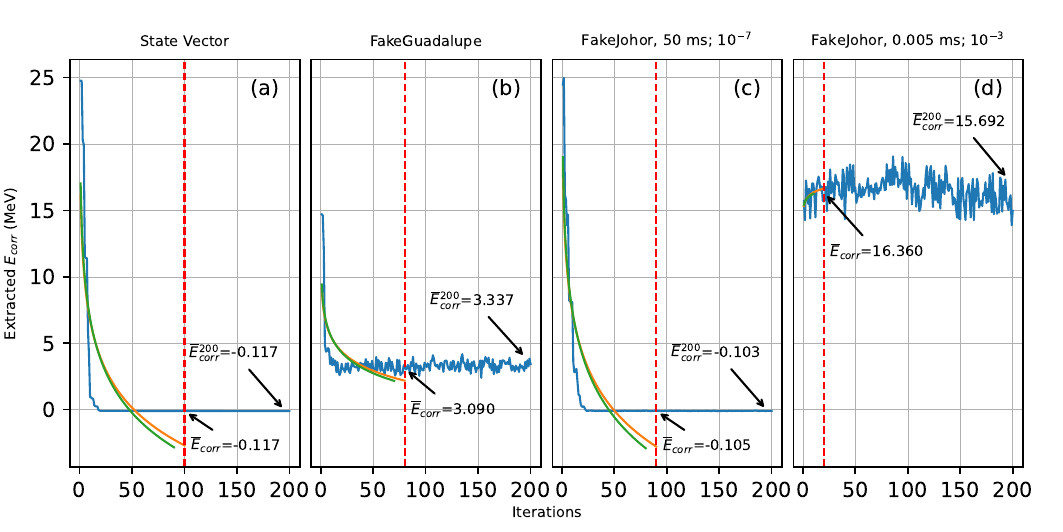}
    \caption{
    Extracted correlation energy $E_{corr}$ as a function of iterative number. Calculations performed on
    (a) state-vector simulator, (b) FakeGuadalupe, (c) FakeJohor with coherence times of $50$ ms and quantum errors of $10^{-7}$ ms and (d) FakeJohor with $0.005$ ms and $10^{-3}$.
    On each subplot, averaged correlation energy over the past 10 iterations are annotated at termination ($\bar{E}_{corr}$) and at the end of the maximum 200 iterations ($\bar{E}_{corr}^{200}$).
    }
    \label{fig:State-vector vs Noisy Simulator} 
\end{figure*}

Figure~\ref{fig:State-vector vs Noisy Simulator} shows the evaluation
of the proposed termination criteria on a state-vector simulator and three different quantum computers specifications.
The evolution of extracted $E_{corr}$ as a function of iteration is shown in Figure~\ref{fig:State-vector vs Noisy Simulator}(a)
for the state-vector simulator.
In the same plot, two lines (in orange and green) shows the last two fitted logarithmic function given in Equation~\eqref{eq: fitted log line}.
Using the termination criteria, 
the state-vector calculations were terminated successfully at the $k = 100$
iteration (indicated with the vertical dashed line).
The averaged $E_{corr}$ over the last 10 iterations obtained was $\bar{E}_{corr} = -0.117$~MeV
reproducing exactly the $\bar{E}_{corr}$ value obtained at maximum $k=200$ iterations (denoted as $\bar{E}_{corr}^{200}$),
reflecting the excellent performance of the termination conditions employed herein.

We also show the good performance of the termination criteria for
noisy quantum simulator as shown in Figure~\ref{fig:State-vector vs Noisy Simulator}(b) 
for the FakeGuadalupe,
and Figure~\ref{fig:State-vector vs Noisy Simulator}(c) for our FakeJohor with coherence times of $50$ ms and quantum errors of $10^{-7}$ ms.
Despite a noisier simulator, the $\bar{E}_{corr}$ values obtained using the termination criteria 
were rather close to the $\bar{E}_{corr}^{200}$
for the respective quantum simulator. 

For extremely noisy simulators e.g.~with coherence times of 0.005~ms and quantum errors $\geq 10^{-3}$, we frequently encountered situations where the calculations did not result in a minimization trend in the evolution of $E_{corr}$ with iteration number.
An example of such calculations is shown in Figure~\ref{fig:State-vector vs Noisy Simulator}(d). In such cases, no convergence is achieved even at maximum iteration.
Using our termination criteria, we successfully terminate the calculations at a much earlier iteration.

\section{The performance of FakeJohors}\label{sec: fakeJohors performance}
We construct the FakeJohors of several specifications, with coherence times ranging from $0.005$~ms to $500$~ms 
and with quantum errors ranging from $10^{-8}$ to $10^{-2}$.
For each specification of the modified quantum simulator, we perform five sets of calculations 
and post-processing was done to select and average the three values closest to the state-vector result.

The performances of different quantum computer specifications to reproduce the correlation energy are plotted as a heatmap in Figure~\ref{fig:heatmap}, where green hues reflect results which are in good agreement with state-vector calculations.
Conversely, red hues signify huge deviations which may exceed the upper bound limit of 0.1~MeV.
On the upper left of the heatmap in Figure~\ref{fig:heatmap}, we see that long coherence times and low quantum errors produce good results as compared to the bottom (short coherence times) and to the right-side (high quantum errors) of the heatmap.
FakeJohors with desirable specifications are bounded by a black dashed line in the heatmap.

We find that a minimum coherence times of 5~ms and a maximum quantum errors of $10^{-4}$ are necessary to reproduce the state-vector results, obtained for a first-order trotterized UpCCD ansatz with circuit depth of 250 within VQE algorithm (see Methods~\ref{subsec: circuit construction}).
To our best knowledge, the coherence times of current transmon qubits are in the range of 0.3~ms to 0.6~ms~\cite{Wang2022, Bal2024}, while the quantum gate errors of $\sim{10^{-5}}$ have been reported for one-qubit gates~\cite{Li2023} and $\sim{10^{-4}}$ for two-qubit gates~\cite{Kubo2024-te}.

On the same heatmap, we also show the overall standard deviations 
of the results as black or white numbers in each pixel of the heatmap.
The overall standard deviations are calculated from standard deviations of the three selected $\bar{E}_{corr}$.
For the pixels within the black dashed line, we observe consistent small standard deviations.
On the other hand, \textit{just outside} the black dashed line, despite the small standard deviations, the performances of FakeJohors deviates significantly from the state-vector's result, $E_{corr}^{SV}$.
Toward small coherence times and large quantum errors, we see increasing standard deviations, peaking at the worst coherence times and highest quantum errors.

\section{Errors from Quantum Computer and Optimization Process}\label{sec: opt vs meas}
The overall error in the final computed $E_{corr}$ is contributed by both the quantum computer
and the optimization process.
For quantum computers, the measurement process inherently yields a spread. 
For a given number of measurement shots, a noisy quantum computer will have a larger spread compared to an ideal quantum computer. 
The optimization process in VQE relies on the measured expectation value to estimate the parameters for the next iteration.
Expectation value with a large spread cause sub-optimal parameter updates, and therefore 
give rise to larger discrepancies at the final iteration.

\begin{figure*}
    \centering
    \includegraphics[width=\linewidth]{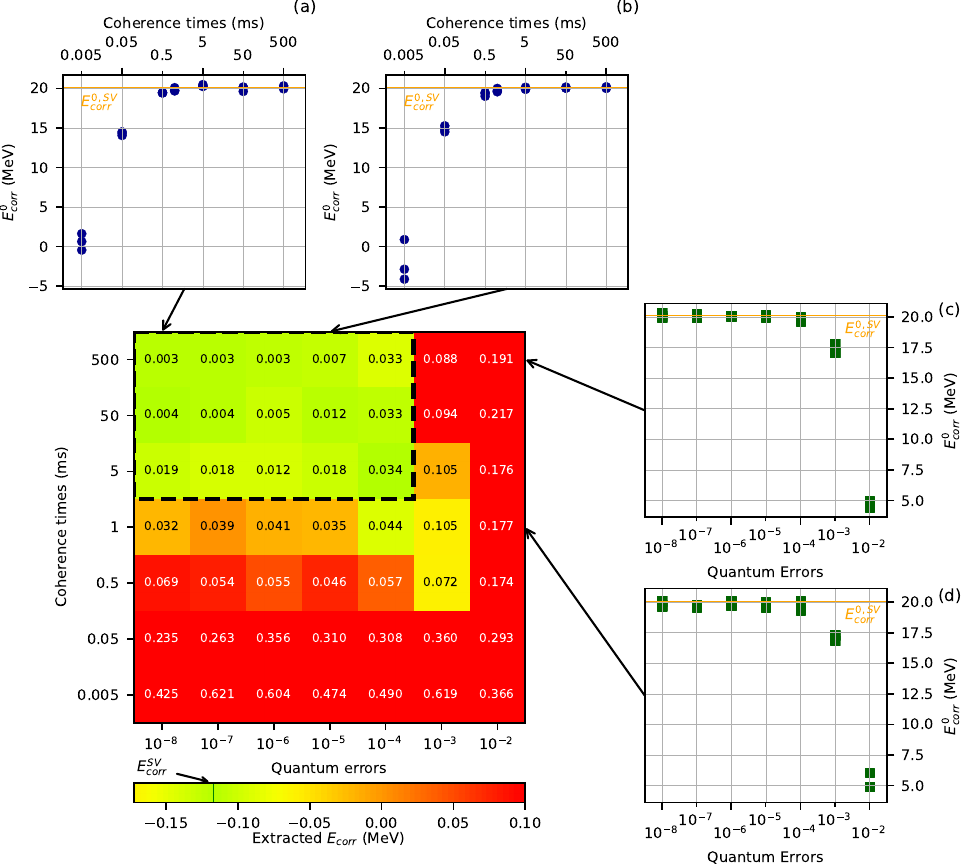}
    \caption{Extracted $\bar{E}_{corr}$ from repeated calculations using simulated quantum computers. The heatmap shows the mean energy from three best results from simulated quantum computers, where green hues reflect values closer to state-vector's result, $E_{corr}^{SV}$, shown as a green vertical line on the color bar. 
    The yellow to red hues signify values that deviate from state-vector results. Some of the averaged $E_{corr}$ which may exceed the upper bound limit of 0.1~MeV, are colored red.
    The number in each box represents the standard deviation of the repeated calculations in unit of MeV.
    (a)-(d) Extracted $E_{corr}$ with respect to the ansatz at iteration 0 as a function of coherence times~(a and b) and quantum errors (c and d).
    Orange line in each of these subplots shows the state-vector evaluated value with the same setting.}
    \label{fig:heatmap}
\end{figure*}

In our results (Figure~\ref{fig:heatmap}), FakeJohors outside the bounded region have worse specifications, and therefore may introduce more errors into the optimization process.
After the final iteration of the optimization process, it becomes impossible to decouple the errors contributed by the quantum computer from the overall performance of the optimization.

In an attempt to identify the errors contributed by the quantum computer, we evaluate the $E_{corr}$ with respect to the ansatz at $k=0$ iteration, hereby denoted as $E^{0}_{corr}$, which did not go through the optimization process.
We then repeated the measurement of $E^{0}_{corr}$ with the ansatz of the same setting, to obtain multiple evaluations of $E^{0}_{corr}$.

For a given quantum errors and for coherence times at 1~ms and below, $E^{0}_{corr}$ deviates further from the $E^{0,SV}_{corr}$, as shown in Figure~\ref{fig:heatmap}(a) and \ref{fig:heatmap}(b).
Likewise, the effect of quantum errors on the discrepancies shows up at $10^{-3}$ and above, as seen in Figure~\ref{fig:heatmap}(c) and \ref{fig:heatmap}(d).
The deviation from $E^{0,SV}_{corr}$, without the SPSA optimization, is a clear indication of errors originating from a noisy quantum computer.

Noise, inherent to NISQ quantum computers, can be modeled in various ways.
As an active field of research, various studies on quantum error modelling aim to understand the impact of quantum errors on quantum algorithms, such as described in Ref.~\cite{Saxena2024}.
For simplicity, within our work, we assume that the errors associated with readout, SPAM and quantum gates incur the same fixed error.
As both the coherence time and quantum errors limit the circuit depth before excessive error is accumulated, strategies to construct shallower circuit (i.e. circuit with fewer gates) are necessary to make use of NISQ quantum computers.
Mapping approaches like block encoding~\cite{Liu2024-um} and treespilation~\cite{Miller2024-pw} have been recently proposed to reduce circuit depth.
Notably, treespilation is claimed to reduce number of CNOT gates by up to 74\%~\cite{Miller2024-pw}.
These pose an opportunity to further reduce the required quantum computer specifications 
for nuclear structure and other similar calculations.

\section{Methods}\label{sec: Method}
\subsection{Quantum circuit construction and evaluation}\label{subsec: circuit construction}
Given a Hamiltonian, $\hat{H}$, and a trial wavefunction ansatz, $\left\vert\psi\right\rangle$, 
VQE algorithm being based on the Rayleigh-Ritz procedure,
attempts to find $\left\vert\psi\right\rangle$ which
minimizes the ground-state energy $E_{\rm VQE}$.
The obtained $E_{\rm VQE}$ is an upper bound value of the actual ground-state energy $E_0$ 
such that
\begin{equation}
    E_{\rm VQE} = \frac{\langle \psi| \hat{H} | \psi \rangle}{\langle \psi  | \psi  \rangle}  \geq E_{0}.
    \label{eq: Ground state energy bound}
\end{equation}

Implementation of the VQE algorithm involves a hybrid quantum-classical computer system for calculations.
Quantum computer is utilized for selected tasks in the whole calculation process namely
for trial wavefunction (ansatz) preparation and measurement of the Hamiltonian expectation value.

To prepare for the ground state, we construct a UpCCD ansatz, which is based on the unitary coupled cluster theory~\cite{Henderson2014, Lee2018} with modification to take only the paired double excitation.
The UpCCD ansatz is defined as 
\begin{equation}
    \left\vert \psi \right\rangle = U\left\vert \Phi _{0}\right\rangle=e^{\hat{T}_2 -\hat{T}_2^\dagger}\left\vert \Phi _{0}\right\rangle, 
    \label{eq:UpCCD ansatz}
\end{equation}
where $U = e^{\hat{T}_2 -\hat{T}_2^\dagger}$ is the unitary operator that prepares the UpCCD ansatz from a Hartree-Fock initial state, $\left\vert \Phi _{0}\right\rangle$, $\hat{T}_{2}$ (and its conjugate $\hat{T}^{\dagger}_2$) is a cluster operator restricted to only pair excitations (and de-excitations). 
The operator $\hat{T}_2$ is expressed in terms of fermionic creation ($\hat{a}^{\dagger}_{k}$) and annihilation ($\hat{a}_{k}$) operators, where $k \in \left\{j,\bar{j},i,\bar{i}\right\}$ as:
\begin{equation}
    \hat{T}_{2}=\sum_{ij}\theta^{j\bar{j}}_{i\bar{i}}\hat{a}^{\dagger}_{j}\hat{a}^{\dagger}_{\bar{j}}\hat{a}_{\bar{i}}\hat{a}_{i},
    \label{eq: double excitation operators}
\end{equation}
where $\theta^{j\bar{j}}_{i\bar{i}}$ is a cluster amplitude. The index $j$ represents an unoccupied state and $i$ represents an occupied state, while $\bar{j}$ and $\bar{i}$ denote their respective conjugates.
Assuming real cluster amplitudes $\theta^{j\bar{j}}_{i\bar{i}} = \theta^{j\bar{j}*}_{i\bar{i}}$, 
the unitary operator in Equation~\eqref{eq:UpCCD ansatz} can be rewritten as
\begin{equation}
U(\vec{\theta}) =\exp \left( \sum_{ij}\theta _{i\bar{\imath}}^{j%
\bar{j}}\hat{\tau}_{i\bar{\imath}}^{j\bar{j}}\right) ,
\label{eq:UpCCD with tau}
\end{equation}%
where
\begin{equation}
    \hat{\tau}_{i\bar{i}}^{j\bar{j}}=\hat{a}_{j}^{\dagger }\hat{a}_{\bar{j}%
    }^{\dagger }\hat{a}_{\bar{\imath}}\hat{a}_{i}-\hat{a}_{i}^{\dagger }\hat{a}_{%
    \bar{i}}^{\dagger }\hat{a}_{\bar{j}}\hat{a}_{j}.
    \label{eq: anti-hermitian excitation operator}
\end{equation}
With the modules from Qiskit (version 0.45.3)~\cite{Qiskit}, the fermion-qubit mapping is done using Jordan-Wigner mapping, given by
\begin{equation}
    \hat{a}_{i}^{\dagger} =\frac{1}{2} \left(\bigotimes\limits_{u=1}^{_{i-1}}Z_{u}\right) \otimes \left( X_{i}-iY_{i} \right),
\end{equation}
\begin{equation}
    \hat{a}_{i} =\frac{1}{2} \left(\bigotimes\limits_{u=1}^{_{i-1}}Z_{u}\right) \otimes \left( X_{i}+iY_{i} \right),
\end{equation}
Equation~\eqref{eq: anti-hermitian excitation operator} is then mapped into
\begin{align}
    \hat{\tau}_{i\bar{i}}^{j\bar{j}} &= \frac{i}{8} \Big( X_j X_{\bar{j}} X_i Y_{\bar{i}} + X_j X_{\bar{j}} Y_i X_{\bar{i}} + Y_j X_{\bar{j}} Y_i Y_{\bar{i}} \notag \\ 
    &\quad + X_j Y_{\bar{j}} Y_i Y_{\bar{i}} - X_j Y_{\bar{j}} X_i X_{\bar{i}} - Y_j X_{\bar{j}} X_i X_{\bar{i}} \notag \\
    &\quad - Y_j Y_{\bar{j}} X_i Y_{\bar{i}} - Y_j Y_{\bar{j}} Y_i X_{\bar{i}} \Big) \bigotimes_{u=\bar{i}+1}^{i-1} Z_u \bigotimes_{v=\bar{j}+1}^{j-1} Z_v \notag \\ 
    &=\frac{i}{8}\left( \sum\limits_{l=1}^{8}\hat{P}_{ij}^{\left(l\right)} \right)
    \bigotimes\limits_{u=\bar{i}+1}^{i-1}Z_{u}\bigotimes\limits_{v=\bar{j}+1}^{j-1}Z_{v}.
    \label{eq:JW mapped excitation operators}
\end{align}
where $\hat{P}_{ij}^{\left(l\right)}$ is a Pauli string (a tensor product of Pauli operators) of length 4, and $l$ sums over all the 8 mapped Pauli strings associated with $i\bar{i}$ to $j\bar{j}$ excitation. It is important to note that Pauli exclusion principle is enforced in the Jordan-Wigner mapping.
Limiting the ansatz to first order trotterized form, the unitary $U$ in Equation~\eqref{eq:UpCCD ansatz} takes the form
\begin{equation}
    U(\vec{\theta})=\prod\limits_{ij}\prod\limits_{l=1}^{8}\exp\left(\frac{i\theta_{i\bar{i}}^{j\bar{j}}}{8}\hat{P}_{ij}^{\left(l\right)}
    \bigotimes\limits_{u=\bar{i}+1}^{i-1}Z_{u}\bigotimes\limits_{v=\bar{j}+1}^{j-1}Z_{v} \right).
    \label{eq:UpCCD quantum circuit operation}
\end{equation}

Using the same mapping, the Hamiltonian in Equation~\eqref{eq:Hamiltonian} is transformed into its qubit equivalent, expressed in terms of Pauli strings $\hat{P}_a \in \left\{ I, X, Y, Z \right\}^{\otimes N}$ for $N$ qubits, given by
\begin{equation}
    \hat{H}_{\text{qubit}} = \sum^{\mathcal{P}}_a \omega_a\hat{P}_a,
    \label{eq:qubit Hamiltonian}
\end{equation}
where $\omega_a$ is the weight for each $\hat{P}_a$, and $\mathcal{P}$ is the total number of $\hat{P}_a$.
In our case, $\hat{P}_a$ has length of $N =12$, corresponding to the number of single-particle states.
Finally, from Equation~\eqref{eq:UpCCD quantum circuit operation} and \eqref{eq:qubit Hamiltonian}, the expectation value of $\hat{H}$, $E$ is given by
\begin{equation}
    E = \sum^{\mathcal{P}}_a \omega_a \left\langle \Phi _{0} \right\vert U^{\dagger}(\vec{\theta}) \hat{P}_a  U(\vec{\theta}) \left\vert \Phi _{0} \right\rangle.
    \label{eq: Energy Expectation value}
\end{equation}

The expectation value in Equation~\eqref{eq: Energy Expectation value} is then transpiled for the targetted simulated quantum computer at the highest optimization level.
Qubit-wise commutative grouping is implemented using Qiskit's module and the expectation value (measurement of the quantum circuits) is then evaluated with 8192 shots.

\subsection{Optimization}
\label{subsec: SPSA optimization} 
The implementation of SPSA involves initial setup of the learning rate given by~\cite{Spall1998}
\begin{equation}
    a_{k}=\frac{a}{\left( A+k+1\right) ^{\alpha }}
\end{equation}
and perturbation value given by 
\begin{equation}
    c_{k}=\frac{c}{\left(k+1\right) ^{\gamma }}
\end{equation}
where $k$ is the optimization step.
The parameters chosen in our work is $\alpha = 0.602$, $\gamma = 0.101$,
$A = 0$, $c = 0.1$ and 
$a$ is calibrated to reduce the expectation value of the first iteration by 1~MeV.

All the calculations started from the same excited state by setting all the initial parameters 
$\theta^{j\bar{j}}_{i\bar{i}}$ to zero except for $\theta^{3\bar{3}}_{1\bar{1}} = 1$ which is chosen arbitrarily, to mitigate barren plateaus as practiced in Ref.~\cite{Kiss2022}.
To ensure proper optimization, the parameter $\theta^{3\bar{3}}_{1\bar{1}}$ has to take on values values other than $\theta^{3\bar{3}}_{1\bar{1}} = \frac{n\pi}{2}$, where $n=0,1,2\dots$. Here, the parameter $\theta^{3\bar{3}}_{1\bar{1}}$ represent the cluster amplitudes in the ansatz~\eqref{eq:UpCCD with tau} that promotes the neutron states from $\left|1\bar{1}\right>$ to $\left|3\bar{3}\right>$ (see Figure~\ref{fig: energy_levels}).

\subsection{FakeJohors Specifications}\label{subsec: FakeJohors spec}
The FakeJohors modified from FakeGuadalupe take the values of coherence times 
such that $T_1 = T_2 = T$ (in milliseconds) with
$$T \in \left\{ 0.005, 0.05, 0.5, 1, 5, 50, 500 \right\}$$
and quantum errors cumulatively referring to the readout error, quantum gate errors, and SPAM errors) with all of them having the same values of
$$\mathcal{E} \in \left\{ 10^{-8}, 10^{-7}, 10^{-6}, 10^{-5}, 10^{-4}, 10^{-3}, 10^{-2} \right\}.$$

For quantum gate errors, we modify the following the one-qubit gates:
\begin{itemize}
    \item Identity,
    \item Rotation Z,
    \item $\sqrt{X}$,
    \item NOT,
\end{itemize}
and the two-qubit gate:
\begin{itemize}
    \item Controlled NOT.
\end{itemize}

\bibliography{ref}

\section{Acknowledgements}
This work is supported by the Universiti Teknologi Malaysia through its UTMShine grant (grant number Q.J130000.2454.09G96).
We would like to extend our sincere gratitude to Dr. Yoon Tiem Leong for the computing resources during the early stage of the work.

\section{Authors contribution}
C.H.W.~performed calculations on quantum computer simulator, visualization, drafting the first draft and editing the manuscript.
M.H.K~performed the HF+BCS calculations, conceptualizing the research framework, reviewing and editing of the manuscript, 
supervision and funding acquisition.
Y.S.Y.~involved in the conceptualizing the research framework, reviewing of the manuscript and supervision.
All authors took part in writing and critical review of the paper.

\end{document}